\documentclass[12pt]{iopart}
\usepackage{graphicx}
\begin{document}

\title[Decoherence and the Branching of Chaos-less Classical Trajectory]{Decoherence and the Branching of Chaos-less Classical Trajectory}

\author{Takuji ISHIKAWA}

%\address{12-3 Midorigaoka, Tsukuba-shi, Ibaraki 305-0863, Japan}
%\ead{st.elmosfire13@gmail.com}
\vspace{10pt}
\begin{indented}
\item[]October 10th 2014
\end{indented}

\begin{abstract}
This time, I found a new rule for decoherence. I used a model without chaos.  As a result, it was shown that not only the intersection of classical trajectories but also branching of classical trajectories are needed for decoherence.  In other words, it was shown that interactions between a main system and environments have to make enough branchings of classical trajectories of the main system for decoherence.
\end{abstract}

%\pacs{03.65.Yz, 03.65.Aa, 03.65.Sq}
%\noindent
\begin{indented}
\item[] {\it Keywords}: decoherence, trajectory, branch
\end{indented}
%\maketitle

\section{Introduction}
 Decoherence is that interference among different quantum states disappear because of dissipation-fluctuation to/from environments$^{1,2}$.
It is said that decoherence makes a quantum system classical.  Because when interference vanishes, an object can not go to other quantum states. It means the object gets classicality.

 I have studied decoherence in a finite system. My motivation was the fact that semi-classical models, such as liquid drop model$^{8}$ or TDHF$^{9}$, are not bad theories in nuclear physics. I wondered the fact. Then I knew Caldeira-Leggett's theory$^{4,5}$. It is said that the key of decoherence is the external environment which is composed with an infinite number of degrees of freedom. Although I was very impressed, I doubted the "infinite numbers". Because I thought that nuclear system, which is finite system, is classical to some extent.

In my previous paper$^{3}$, there is an isolated system made with three degrees of freedom. And it is shown that one selected degree of freedom(particle-1) loses its quantum mechanical property because of the other two degrees of freedom(particle-2 and -3). That is , decoherence occurs for particle-1.

The quantum mechanical model in the paper has a special property that decoherence occurs when classical trajectories intersect in a corresponding classical model. But we can not simply conclude that "Classical crossing makes decoherence". In Caldeira-Leggett's model$^{4,5}$ in which a pair of Gaussian packets is in a harmonic oscillator potential with a heat bath, decoherence does not occur when those Gaussian packets cross in weakly damping case. And also in my model which I will show you in this paper, decoherence won't occur when coupling between main system and environments is not enough. Strictly speaking, intersections of classical trajectories which contain some condition will make dephasing in corresponding quantum mechanical system, and the dephasing makes decoherence. That condition will be shown in this paper.

This time, I used a model which added an external harmonic oscillator potential which interacts with only a main degree of freedom(particle-1) to my previous model which composed with three particles being tied with springs each other.  As a result, it was shown that not only the intersection of classical trajectories but also branchings of classical trajectories are needed for decoherence.  In other words, it was shown that interactions between a main system and environments have to make enough branchings of classical trajectories of the main system for decoherence.

\section{Methods}
Please imagine a closed system, there three particles are mutually tied with springs which have different angular frequencies respectively. Lagrangian \(L\) is as follow. (\(x_1 , m_1\) means position and mass of particle-1 respectively, etc.)
{\small
\begin{eqnarray}
 L &=& \frac{m_1}{2} \dot{x_1}^{2} +  \frac{m_2}{2} \dot{x_2}^{2} + \frac{m_3}{2} \dot{x_3}^{2} -\frac{K}{2} x_1^2 \nonumber \\
 &\quad& -\frac{K_{12}}{2} (x_1-x_2)^{2} -\frac{K_{23}}{2} (x_2-x_3)^{2} \nonumber \\
 &\quad& \quad -\frac{K_{31}}{2} (x_3-x_1)^{2}  \label{e1}
\end{eqnarray}
}
 %For this model, I will apply the technique of Caldeira-Leggett. This three particles model is a extreme reduction of their ''harmonic oscillator plus reservoir model''.  

 For this Lagrangian Eq.(\( \ref{e1}  \)), we can get Feynman propagator by a way of change of variables as I did in my previous paper. But this Lagrangian is different from a Lagrangian of my previous model$^{3}$. Because, in this new model, respective masses can be  different each other , and there is an external harmonic potential for particle-1. In this case, it seems to be difficult to get a diagonal matrix and an orthogonal matrix to change of variables analｙtically. Therefore, this time, I used numerical solution for them. I got the program to derive those matrices from ``Miso no Keisan Buturigaku(http://www.geocities.jp/supermisosan/)''. Using the propagator, we can write a time evolution of wave function of three body system.

Initial wave function of the three body system is the product of wave functions of each particles at initial time $t_0$. And the each initial state is Schr\"odinger cat state,
{\small
\begin{eqnarray}
 &&\psi_1(x_{1(0)},t_0) = \tilde{ N_1 } \times \quad\quad \quad \nonumber \\
 \nonumber \\
 &&\left[ \  \exp\left\{-\frac{x_{1(0)}^2}{4\sigma_1^2}\right\}+\exp\left\{ -\frac{(x_{1(0)}-d_1)^2}{4\sigma_1^2}\right\} \ \right] \nonumber \\ \label{e3} 
\end{eqnarray}
}
etc.. Here, \(x_{1(0)}\) means \(x_1(t_0)\), \(\sigma_1\) means half width of packet and \(\tilde{ N_1 }\) means a normalization constant. When we are only interested in the information about a degree of freedom (particle-1) as a subsystem,  we should integrate out the information about particle-2 and -3 as environments. Then we can get the information about particle-1 only, that is, the reduced density function for particle-1, $\tilde{\rho}_1$.
{\small
\begin{eqnarray}
 &&\tilde{\rho}_1^{(reduced)}(x_1,t) = \nonumber \\ \nonumber \\
 &&\quad\quad\quad \int_{-\infty}^{\infty} \int_{-\infty}^{\infty}dx_2 \ dx_3 \ \rho^{(total)}(x_1,x_2,x_3,t) \nonumber \\ \label{e4}
\end{eqnarray}
}
 Furthermore, we can separate the quantum interference term from the reduced density, and a disappearance of the interference means decoherence. I used numerical integration in Eq.(\ref{e4}) and final normalization.

 This procedure above is basically the same as Caldeira-Leggett's technique$^{4,5}$. They used an influence functional method in which a Feynman propagator includes effects of environmental degrees of freedom. The difference between my procedure here and theirs is only an order of integrals and path integrals.

On the other hand, we can draw corresponding classical trajectories.
In a classical harmonic three body problem, we can draw spatial trajectories of particle-1 ($x_1=x_1(t)$) versus time t. Each particle has 2 initial positions which are corresponding to centers of two Gaussian packets in quantum system, and their all initial velocities are set 0. Then we can draw $2^3 = 8$ trajectories on ($x_1$-t) plane.  It is suggested that there should be some relationship between behaviors of classical trajectories and  decoherence in a corresponding quantum system$^{3}$.

\section{Result}

Time evolutions of a reduced density function $\tilde{\rho}_1$ derived from quantum mechanical calculation are shown in graphs (a)-(l) at FIG.\ref{f1} and FIG.\ref{f2}. As you can see, there are 2 packets in each graph. The right packet was at the origin ($x_1=0$) initially. The left packet was at a distance ($x_1=d_1$) initially. And there are 2 wave-like lines in each graph. The lower wave-like line is the quantum interference between these 2 packets. Disappearance of the interference means an emergence of classicality, that is decoherence. Time evolutions of the maximum values of interferences are shown in graphs (n) at FIG.\ref{f1} and FIG.\ref{f2} respectively. The upper wave-like line is the reduced density function for particle-1. On the other hand, corresponding classical trajectories \(x_1(t)\) are shown in graphs (m) at FIG.\ref{f1} and FIG.\ref{f2}.

 Both in two cases below, Lagrangian is Eq.(\( \ref{e1}  \)). Masses of three particles are \(m_1\)=1.5, \(m_2\)=1.0 and \(m_3\)=1.0 respectively.  An external harmonic potential for particle-1 is corresponding to a spring constant \(K\)=2.5. Initial wave function for particle-1 is Eq.(\( \ref{e3}  \)) in which \(d_1\)=-5.0. For particle-2 and particle-3, \(d_2\)=6.0, \(d_3\)=7.5 respectively. And normalization constants \(\tilde{ N_1 },\tilde{ N_2 },\tilde{ N_3 }\) can be set 1, because we can use a numerical normalization. And \( \hbar=1.0\).

\subsection{Weak Coupling Case} Coupling constants among a main degree of freedom (particle-1) and two environmental degrees of freedom (particle-2 and particle-3) are \(K_{12}\)=0.01442, \(K_{31}\)=0.01732. A coupling constant between environmental degrees of freedom particle-2 and particle-3 is \(K_{23}\)=1.02236.

\begin{figure}[tbh]
\includegraphics[scale=.39]{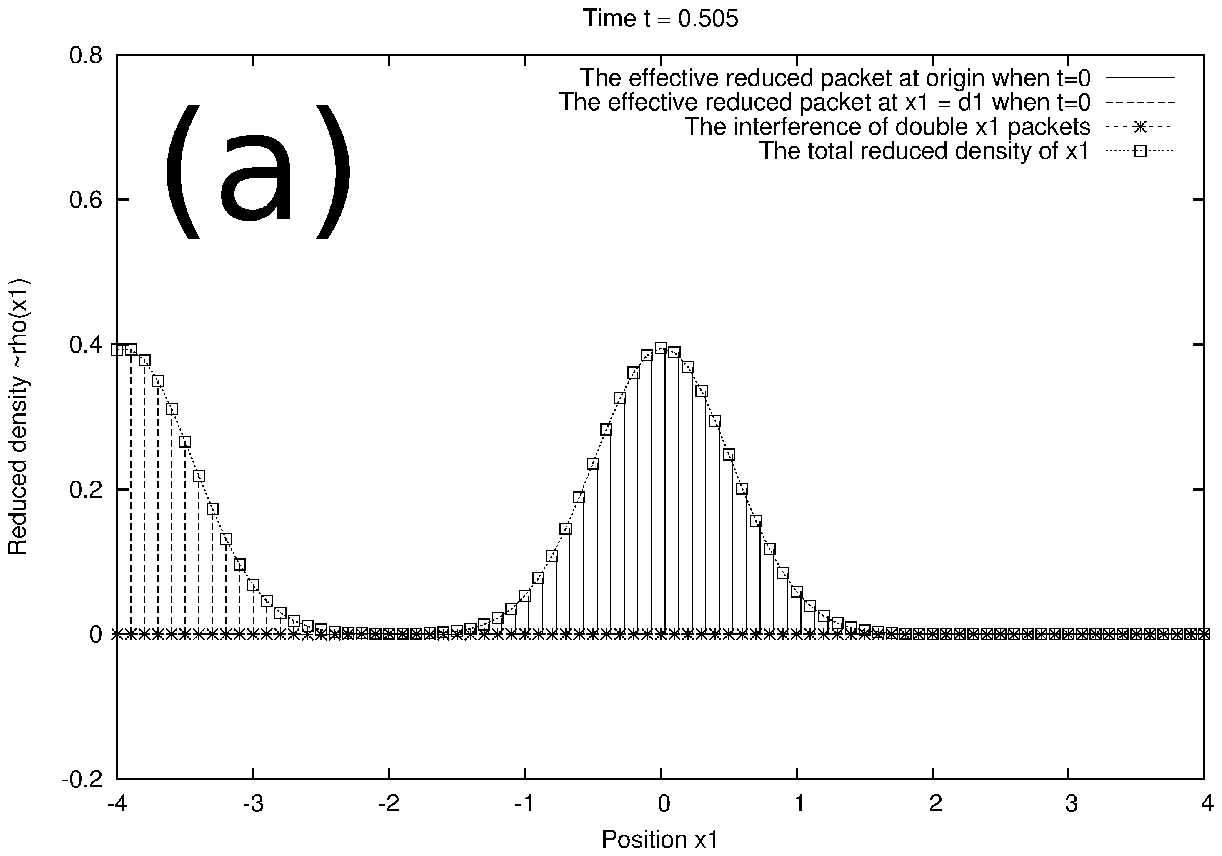}
\includegraphics[scale=.39]{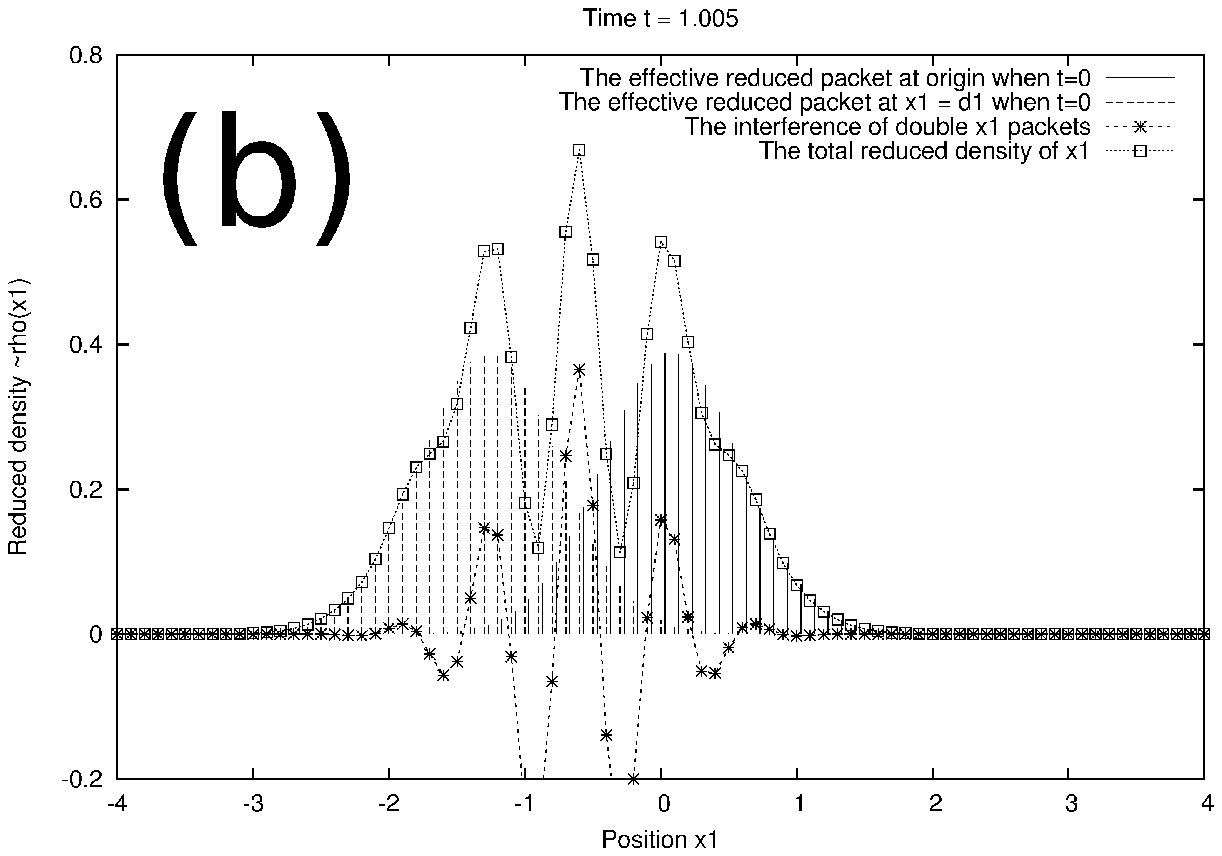}
\includegraphics[scale=.39]{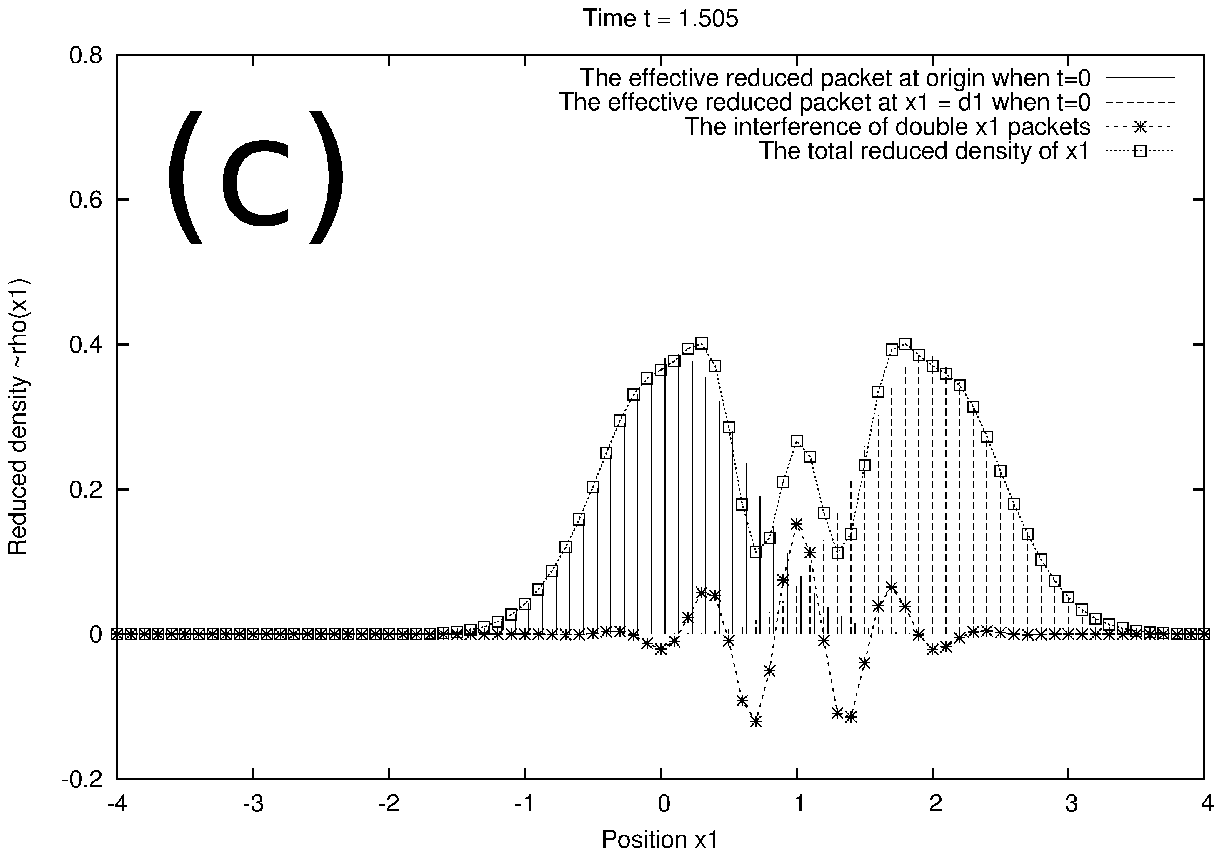}

\includegraphics[scale=.39]{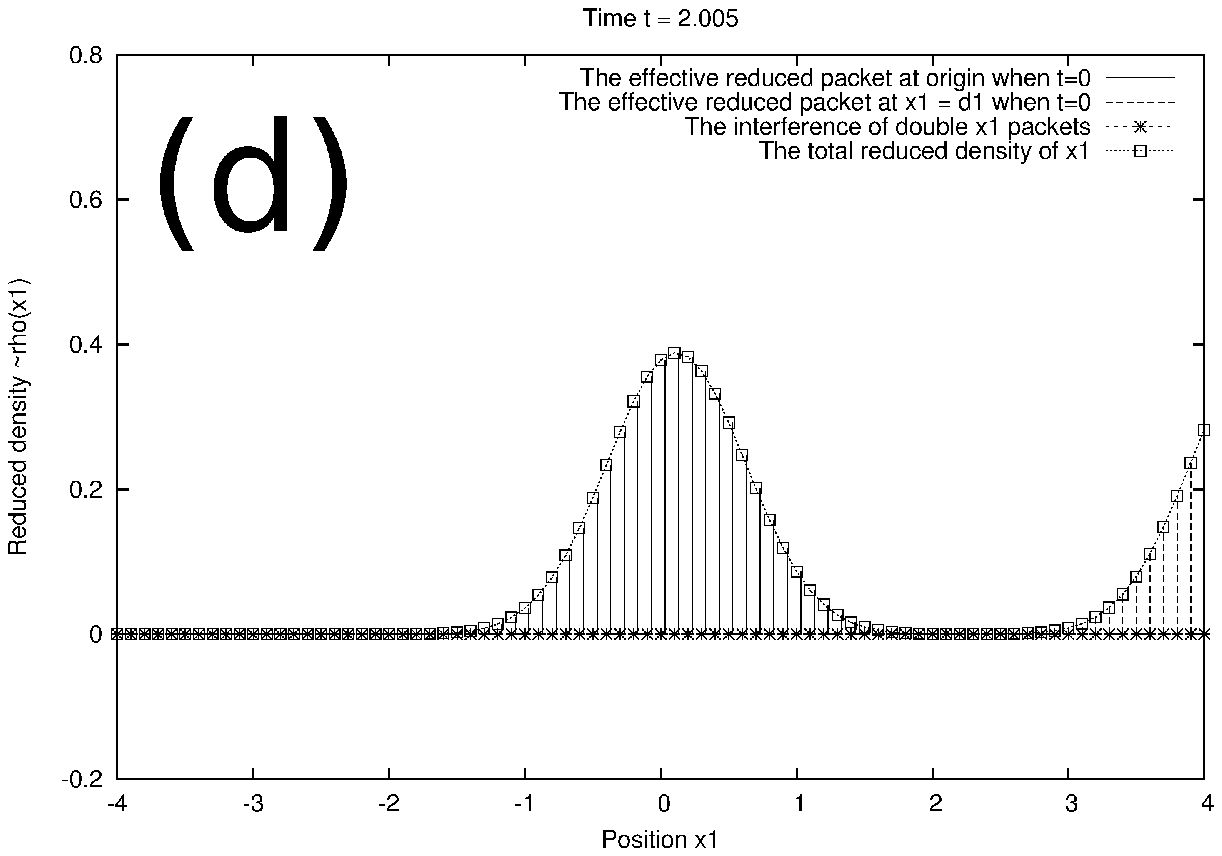}
\includegraphics[scale=.39]{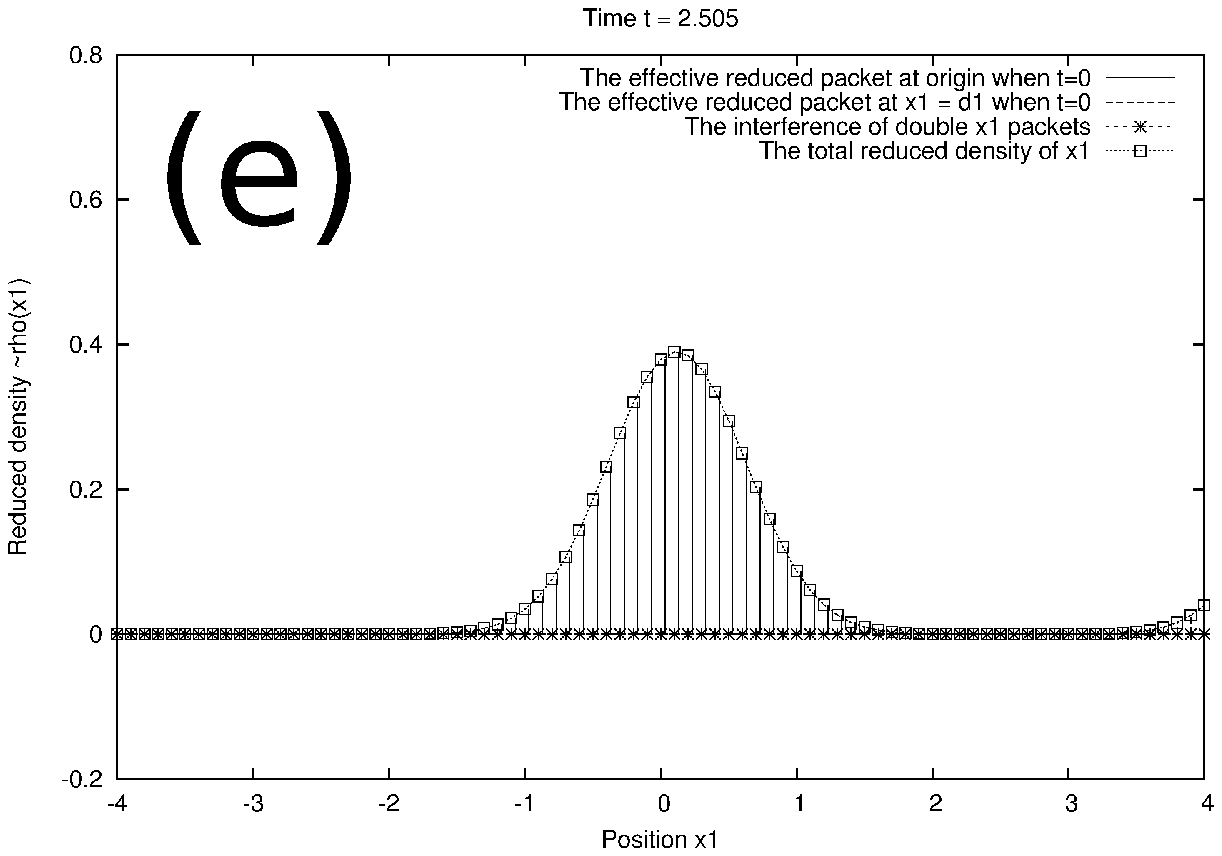}
\includegraphics[scale=.39]{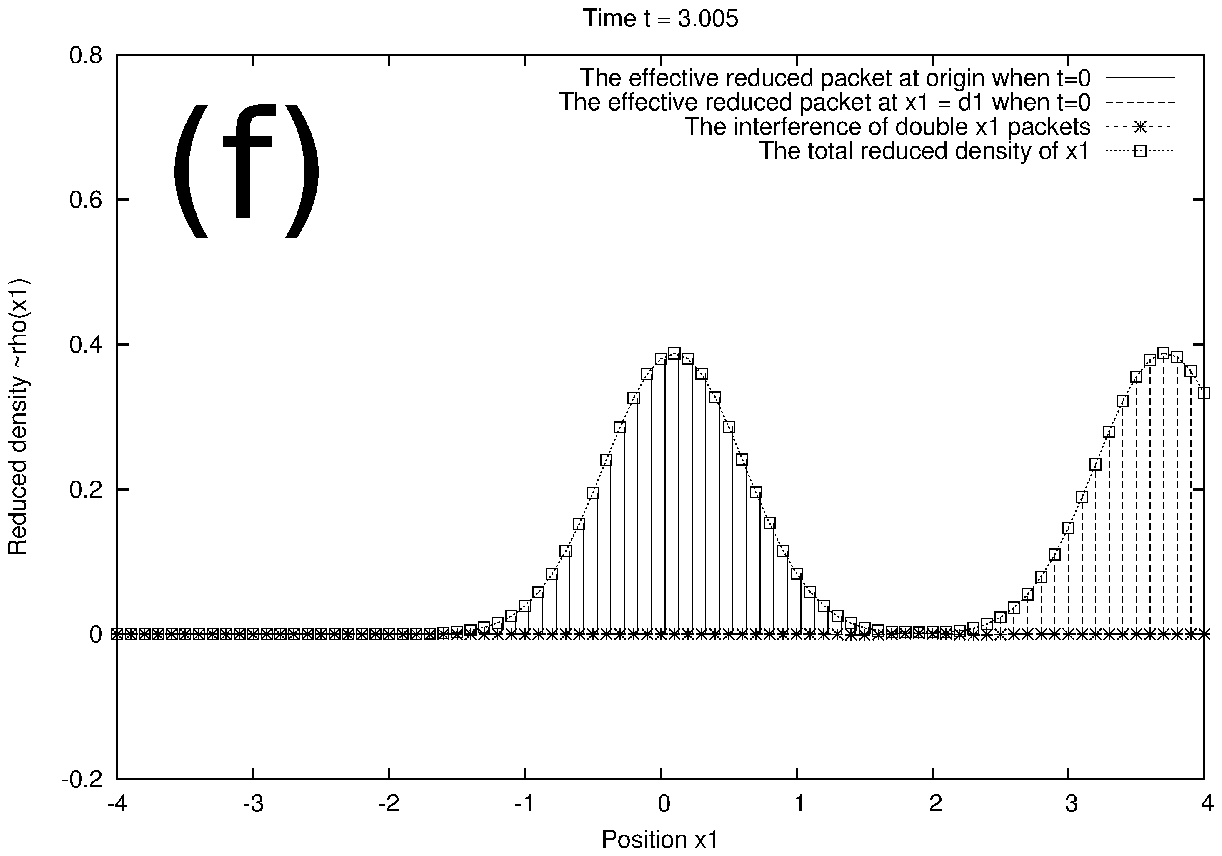}

\includegraphics[scale=.39]{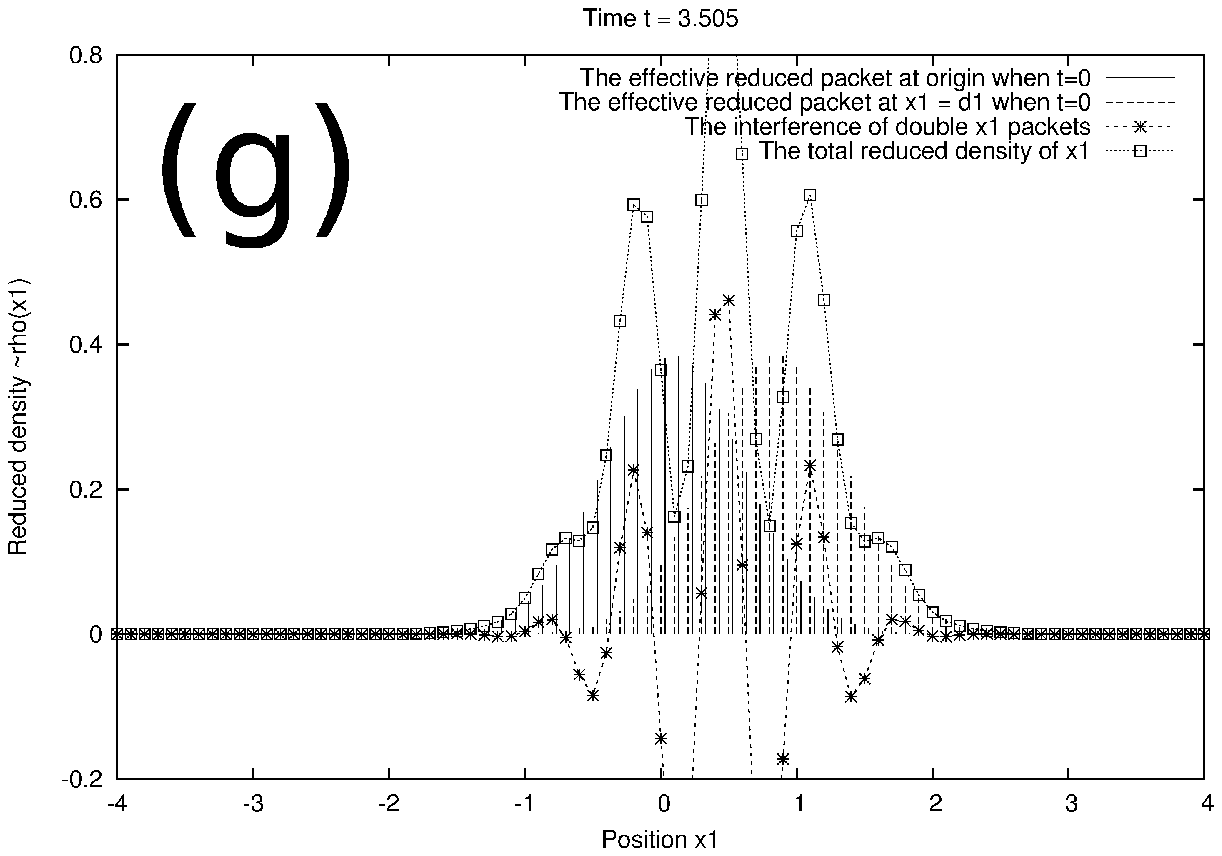}
\includegraphics[scale=.39]{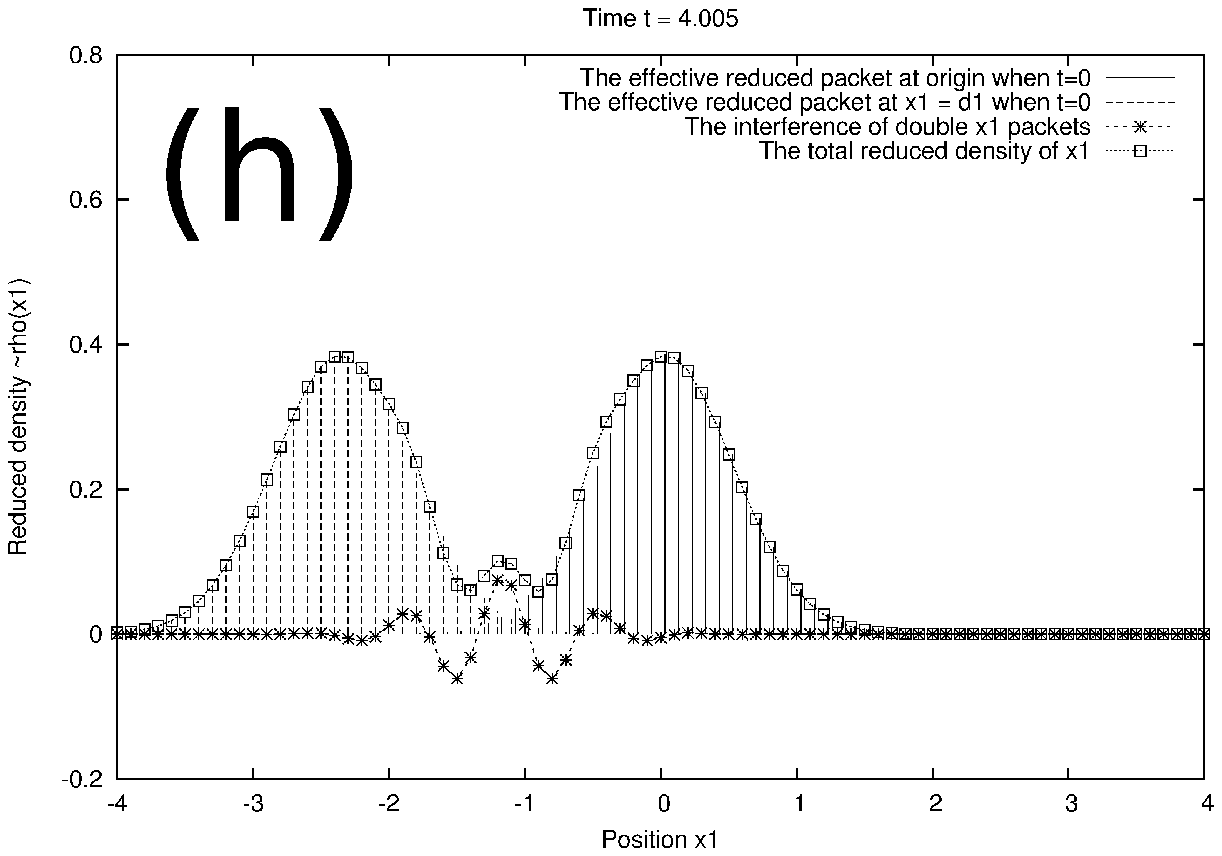}
\includegraphics[scale=.39]{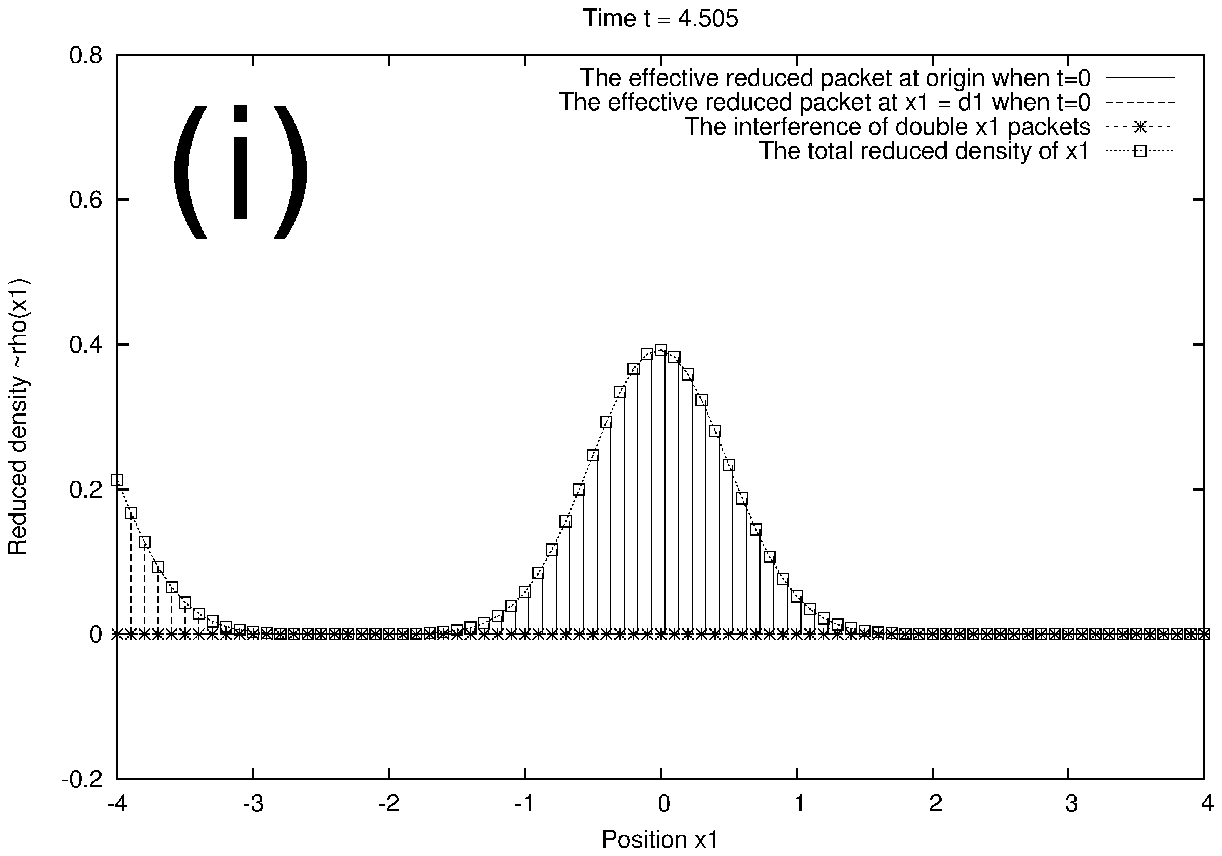}

\includegraphics[scale=.39]{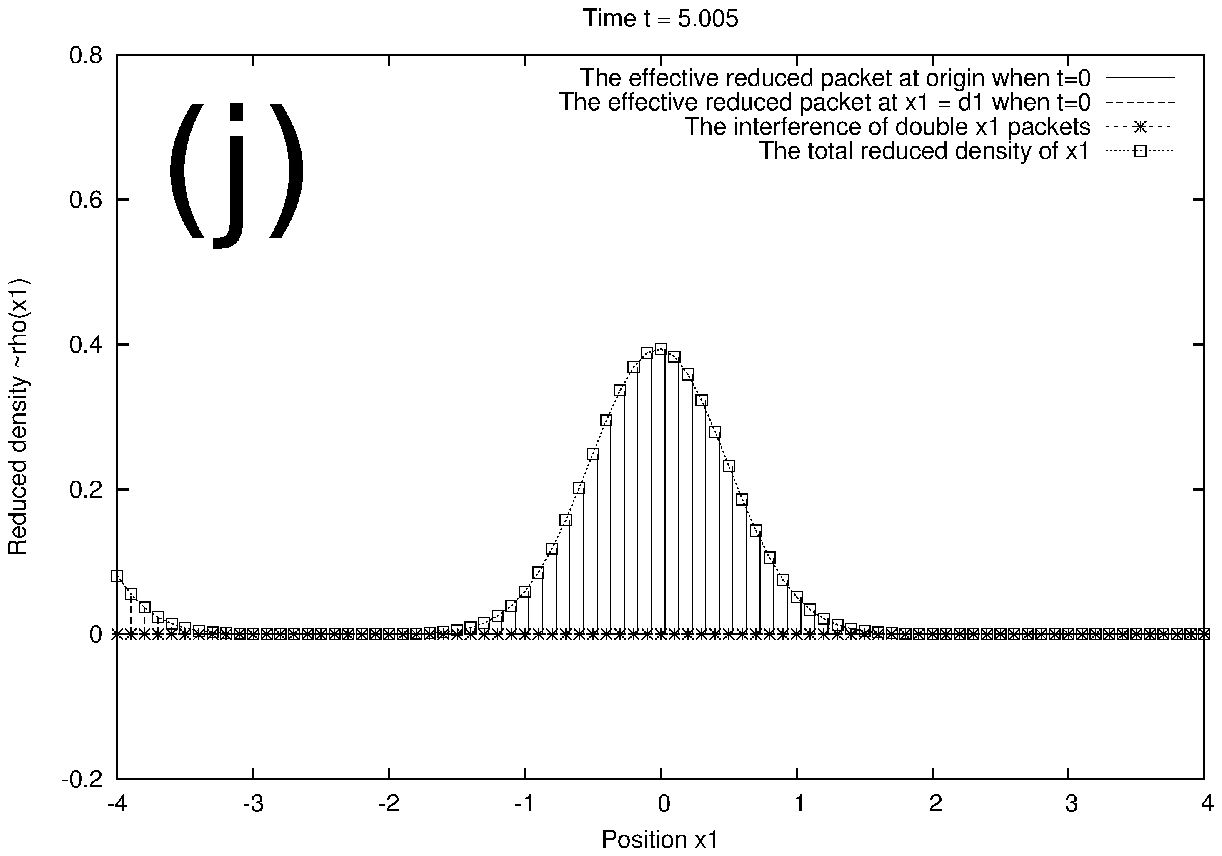}
\includegraphics[scale=.39]{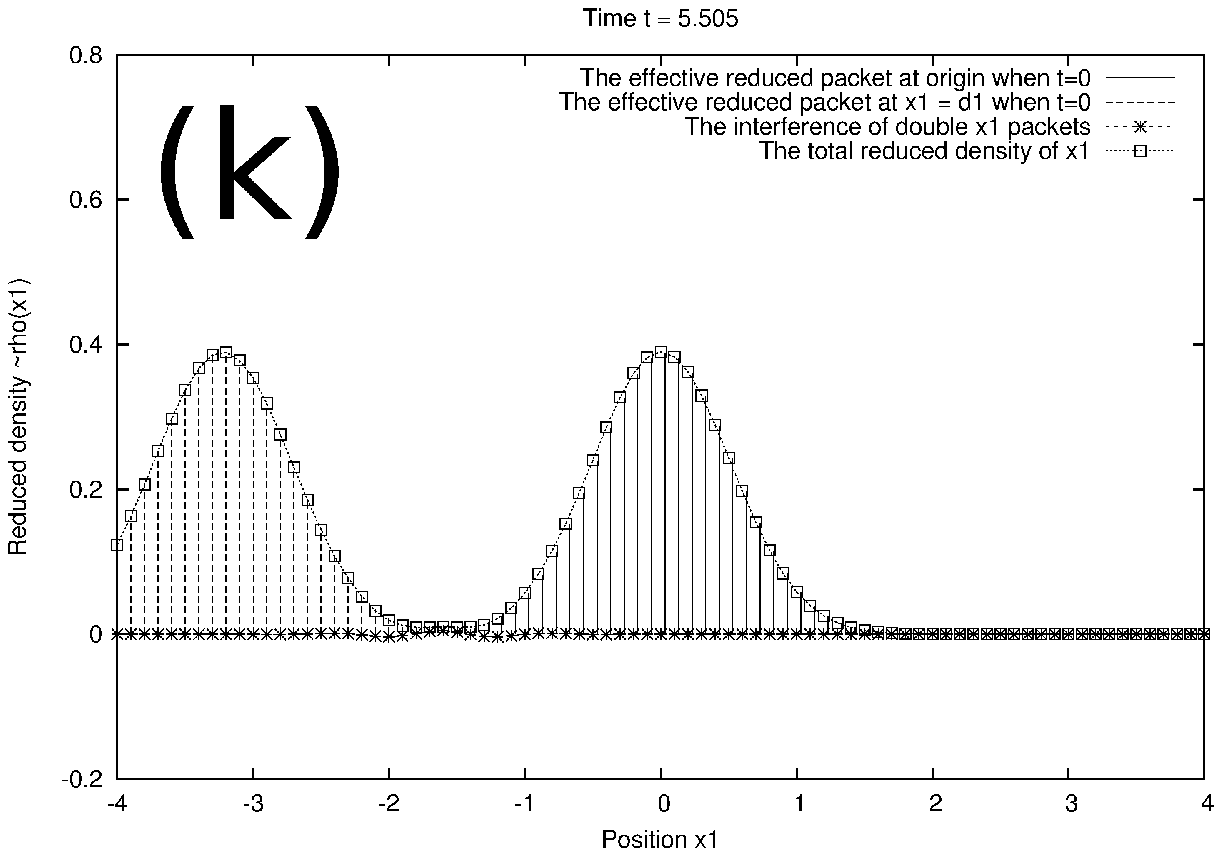}
\includegraphics[scale=.39]{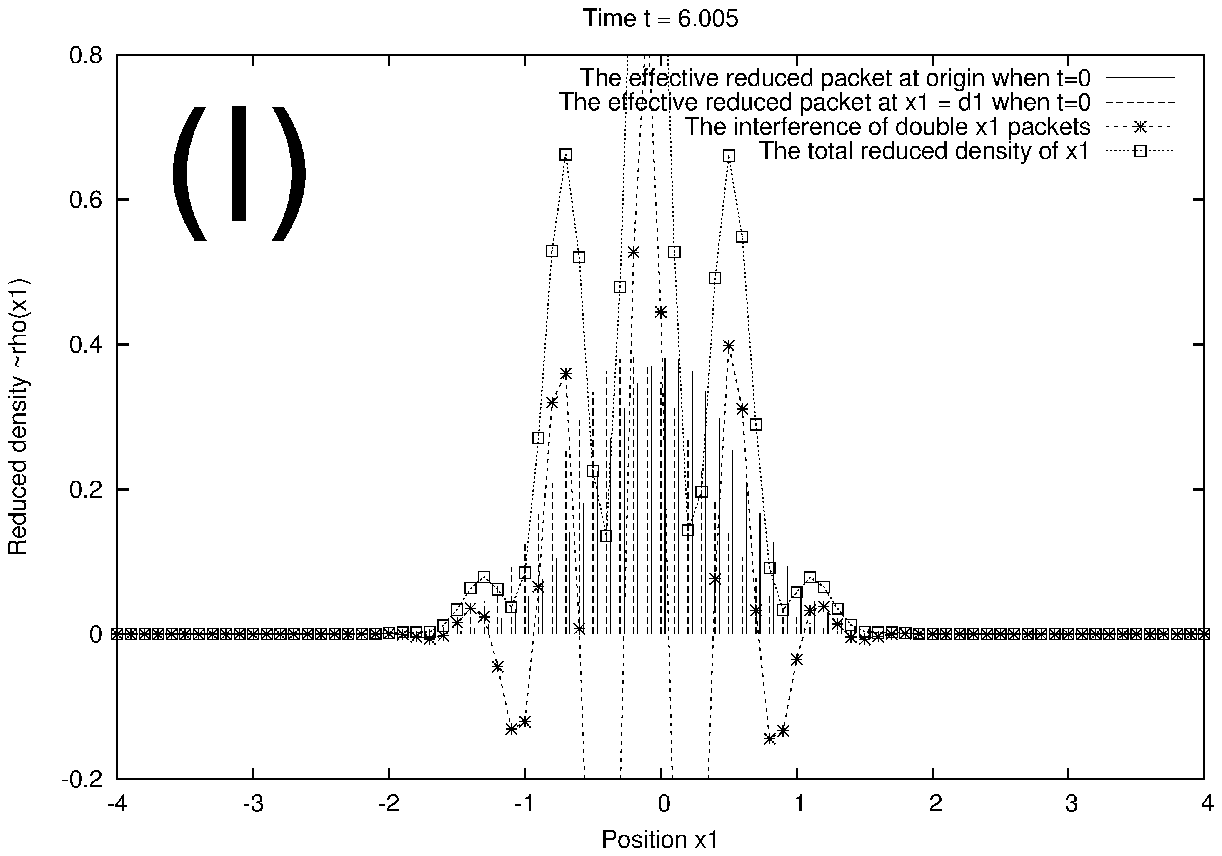}

\includegraphics[scale=.28]{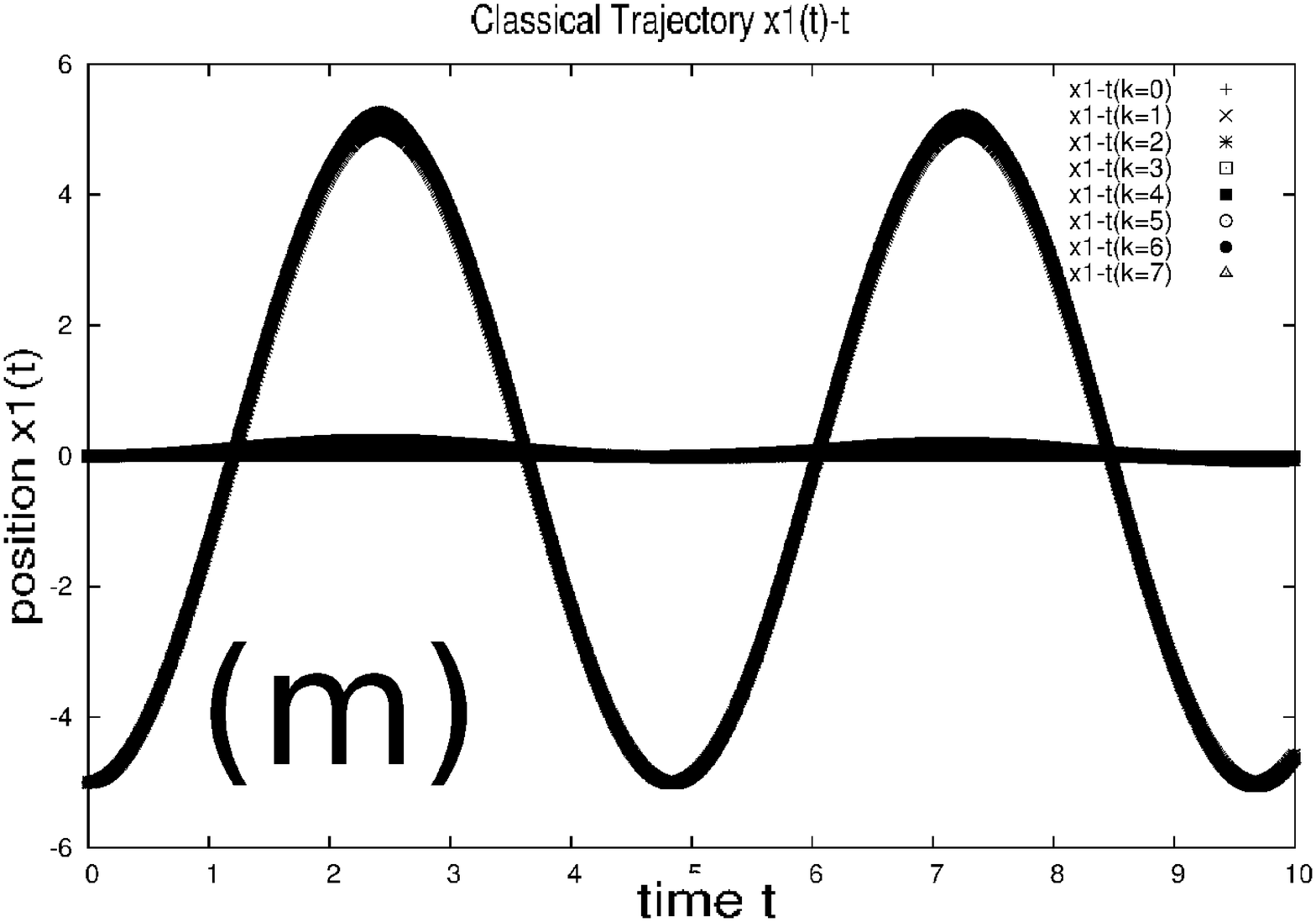}
\includegraphics[scale=.28]{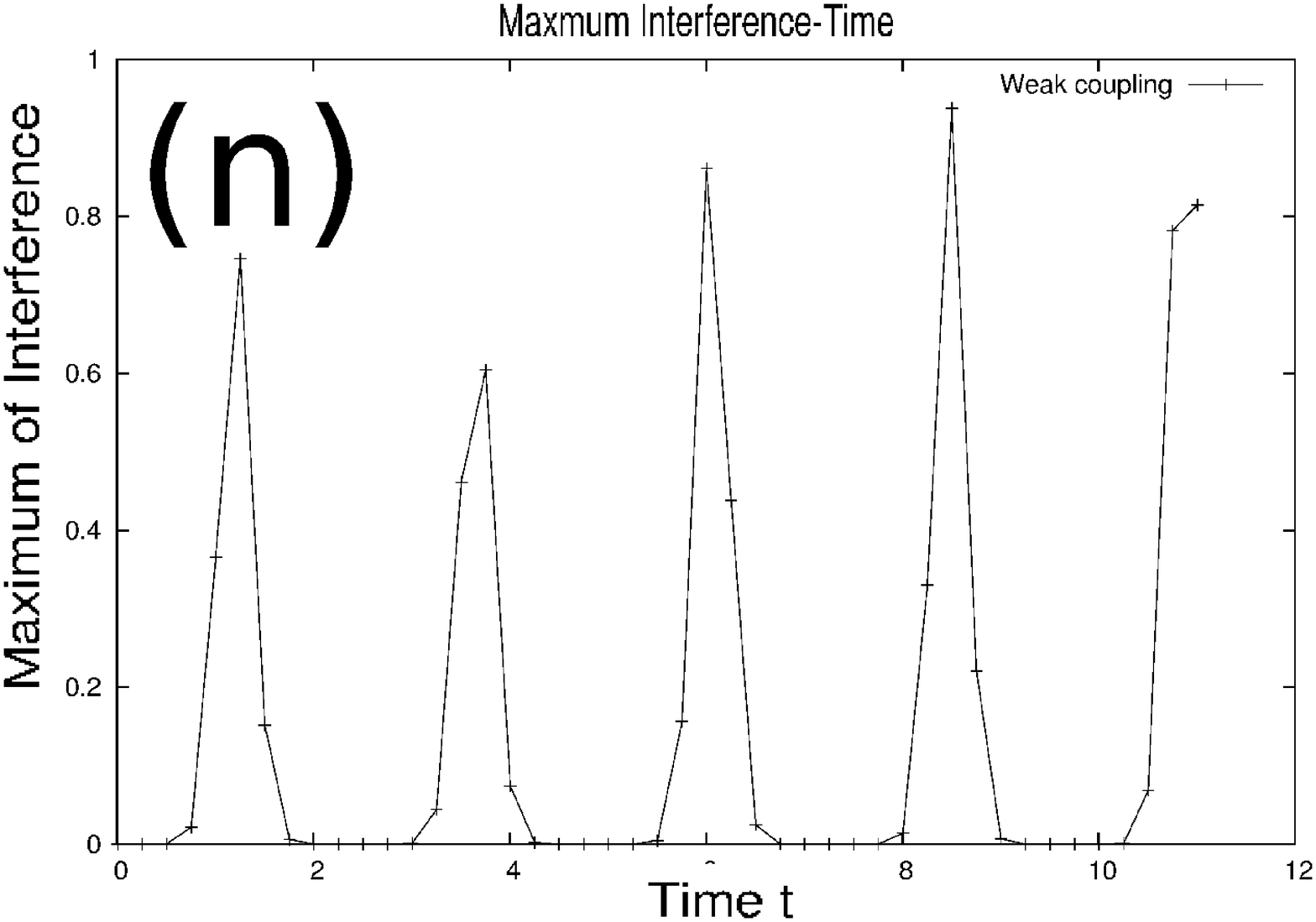}

\caption{Graphs (a)-(l) show a time evolution of the reduced dencity of particle-1 for weak coupling case. (a) Time t=0.505, (b) t=1.005, (c) t=1.505, (d) t=2.005, (e) t=2.505, (f) t=3.005, (g) t=3.505, (h) t=4.005, (i) t=4.505, (j) t=5.005, (k) t=5.505, (l) t=6.005. And (m) shows a time evolution of classical trajectories of particle-1, \(x_1(t)\), (n) shows a time evolution of the maximum value of quantum interference, for weak coupling case.}
\label{f1}
\end{figure}

 In graphs derived from a quantum mechanical calculation((a)-(l) and (n) in FIG.\ref{f1}), when two Gaussian packets cross, interference between them remains to be strong and decoherence does not occur.

 At this moment, how are corresponding classical trajectories? As you can see in a figure of classical trajectories (Image (m) in FIG.\ref{f1}), although eight classical trajectories which are corresponding to respective initial states of three particles cross, they are not strongly branched. Classical trajectories with the same initial position draw a similar curve. It means dephasing for quantum system is not enough.

\subsection{Strong Coupling Case}
 Coupling constants among a main degree of freedom (particle-1) and two environmental degrees of freedom (particle-2 and particle-3) are \(K_{12}\)=0.1442, \(K_{31}\)=0.1732 which are 10 times as much as the weak coupling case above. On the other hand, a coupling constant between environmental degrees of freedom particle-2 and particle-3 is the same as above,\( K_{23}\)=1.02236. 
 
\begin{figure}[tbh]
\includegraphics[scale=.39]{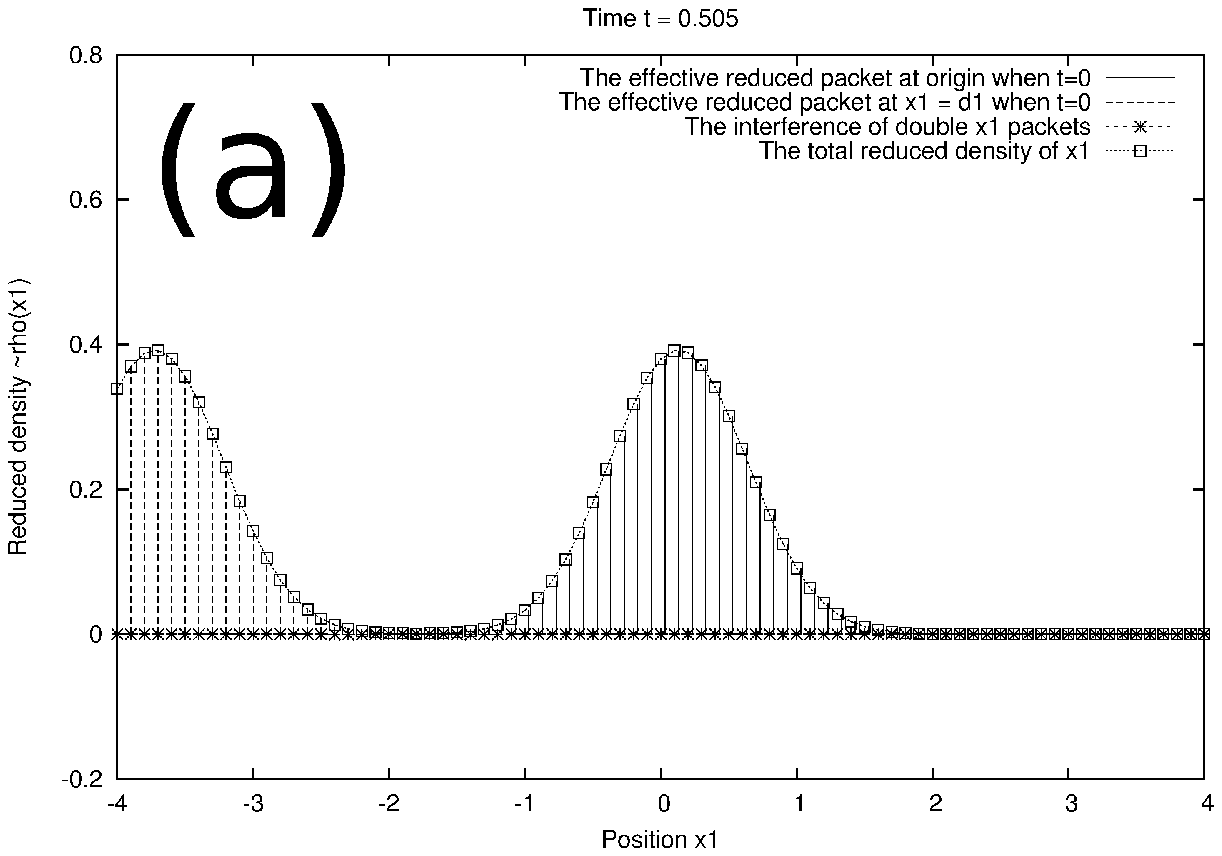}
\includegraphics[scale=.39]{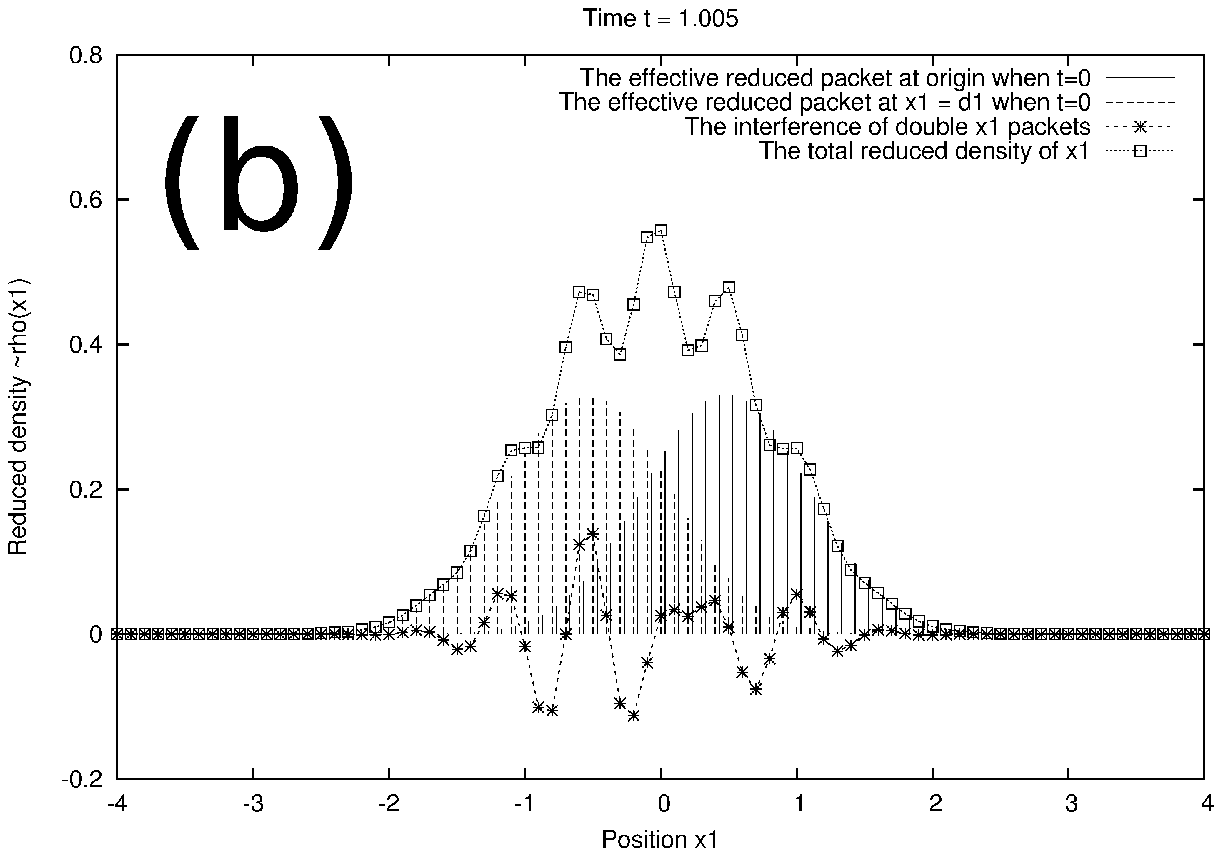}
\includegraphics[scale=.39]{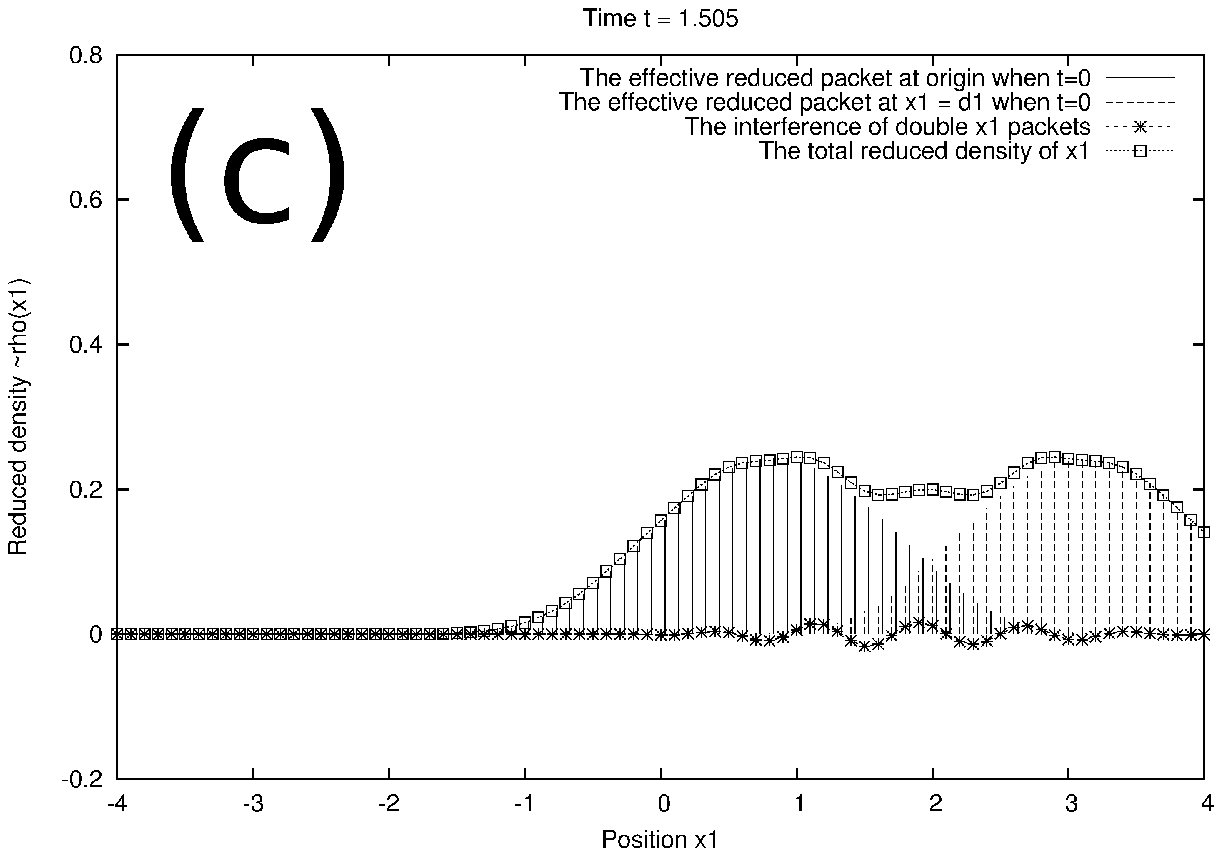}

\includegraphics[scale=.39]{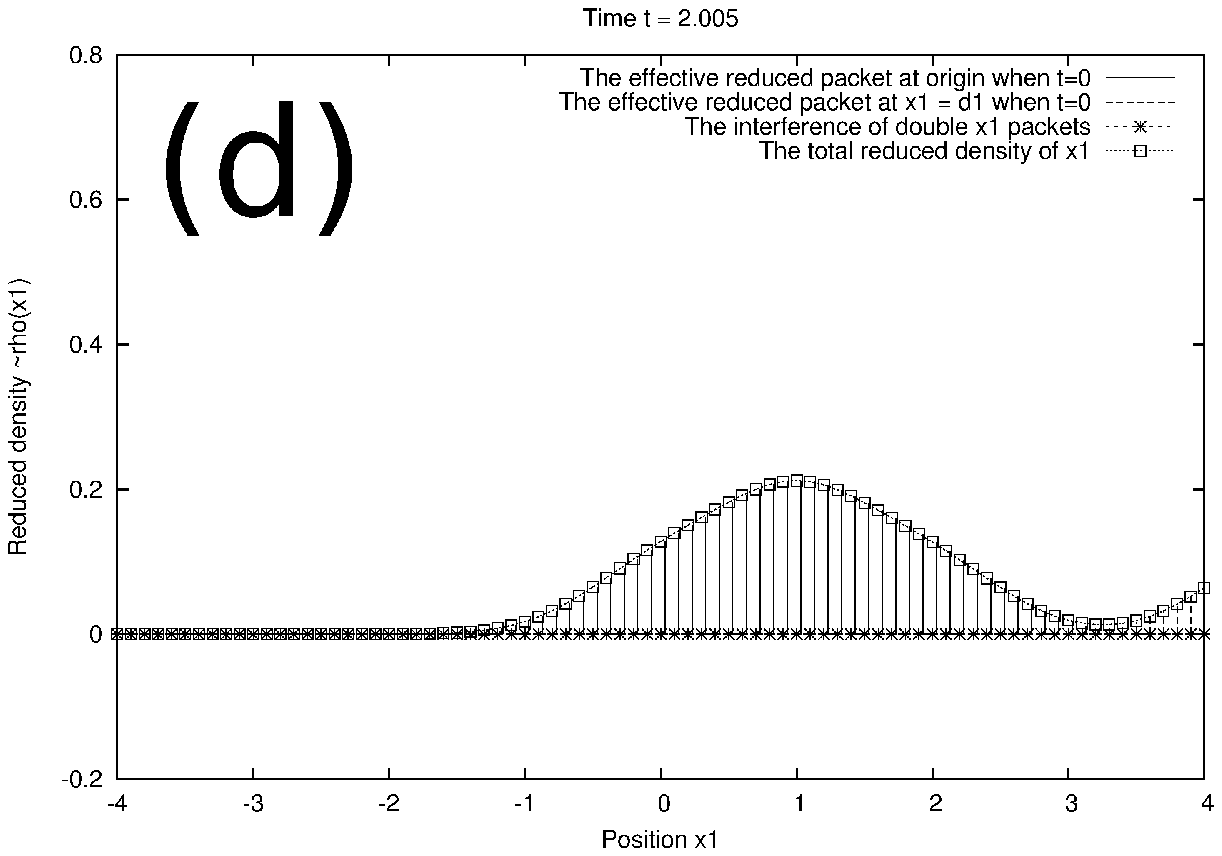}
\includegraphics[scale=.39]{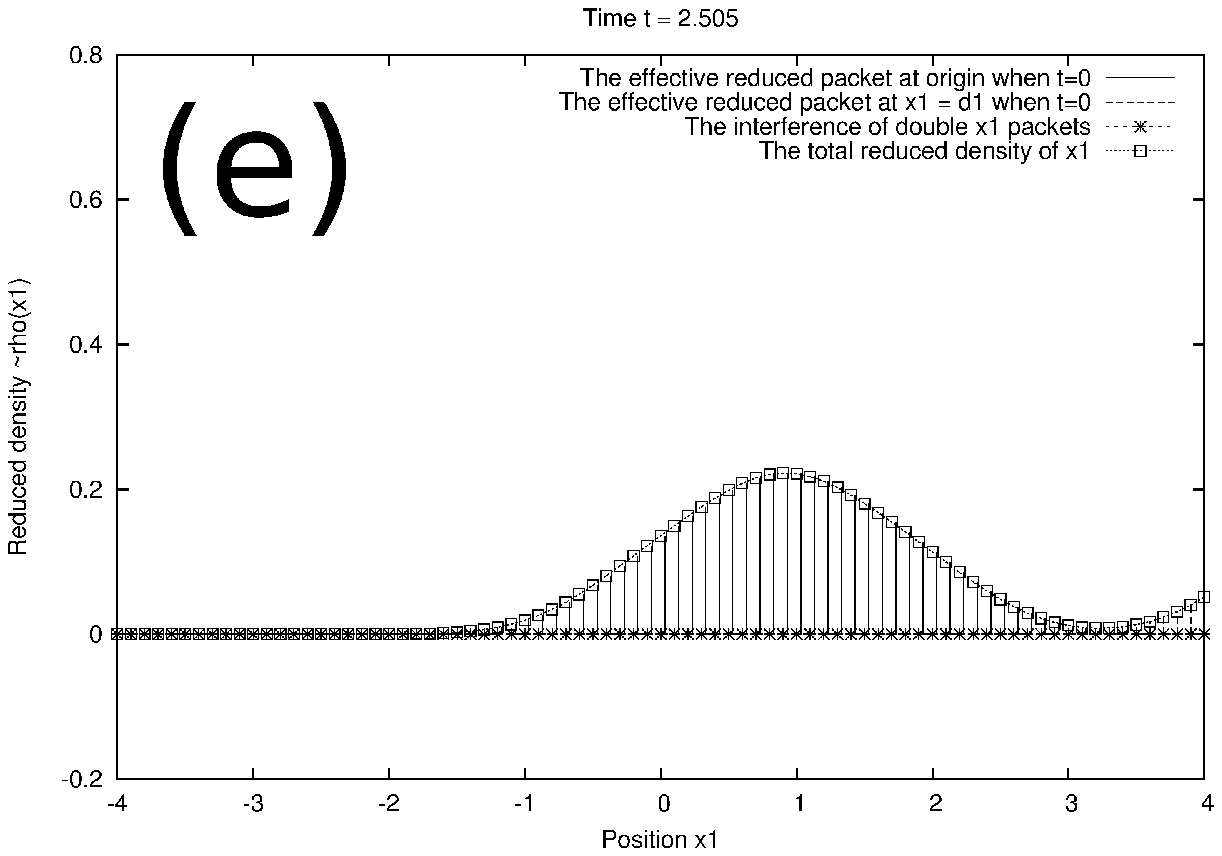}
\includegraphics[scale=.39]{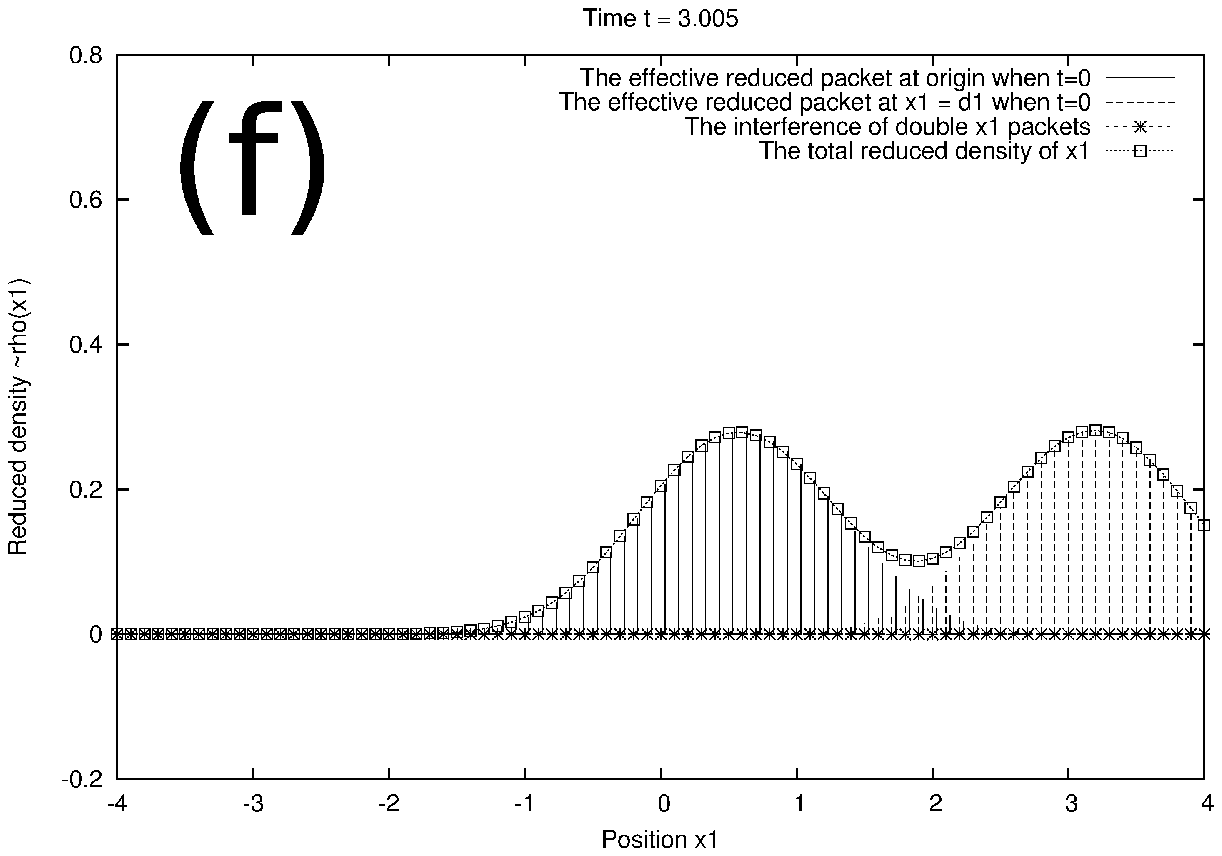}

\includegraphics[scale=.39]{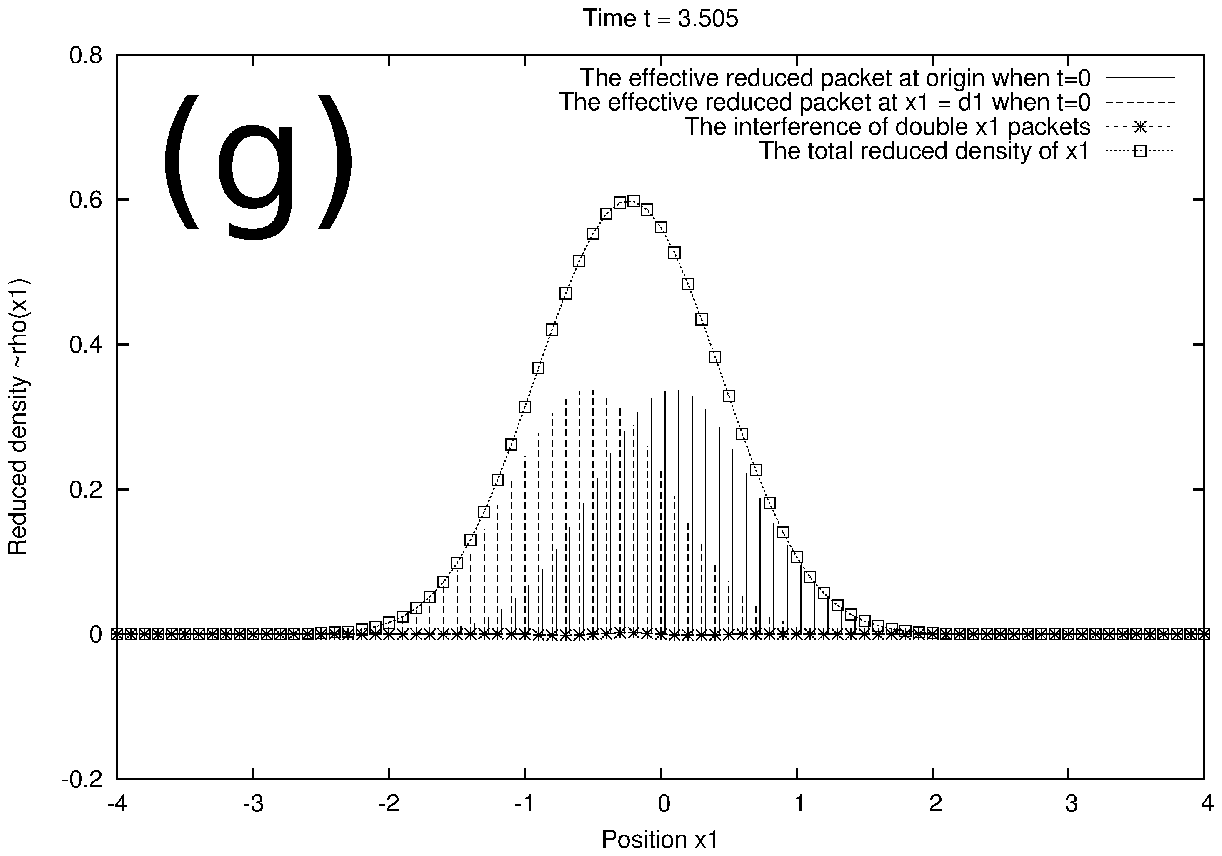}
\includegraphics[scale=.39]{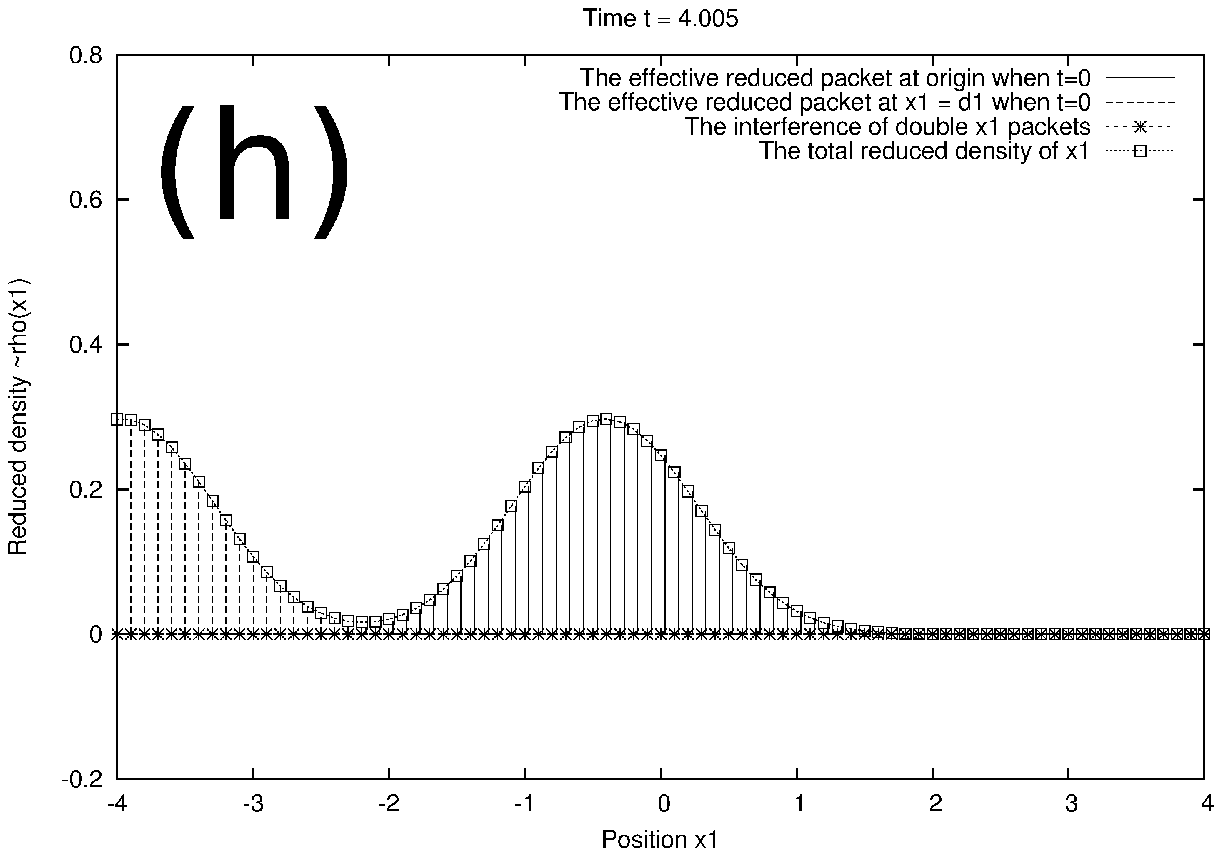}
\includegraphics[scale=.39]{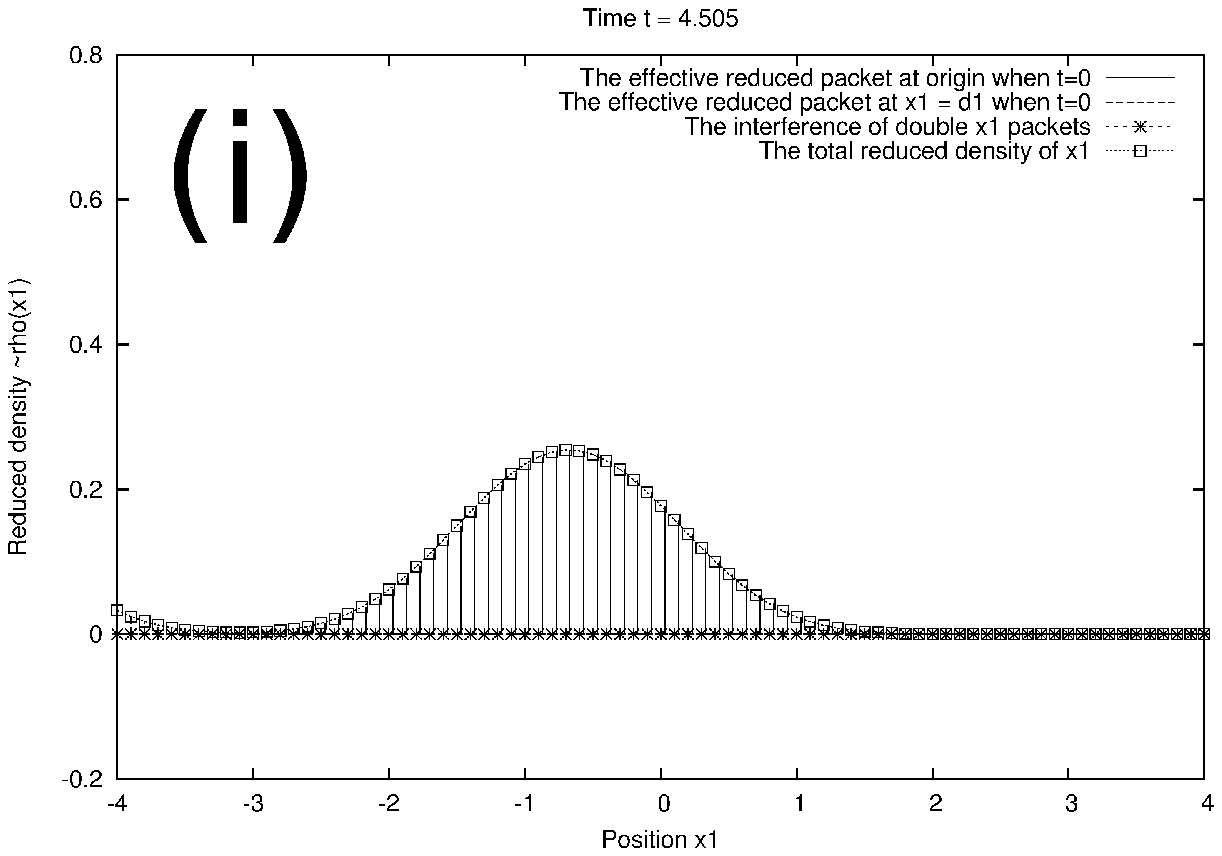}

\includegraphics[scale=.39]{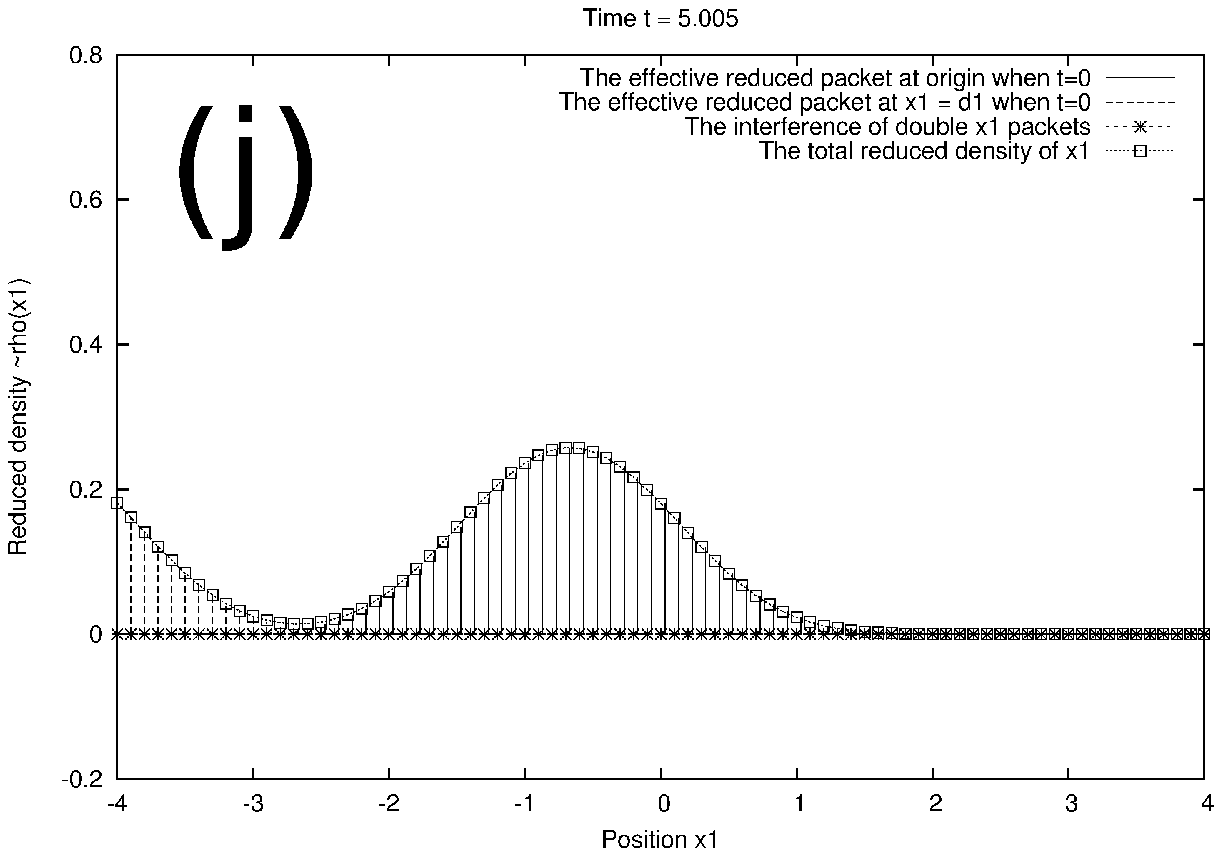}
\includegraphics[scale=.39]{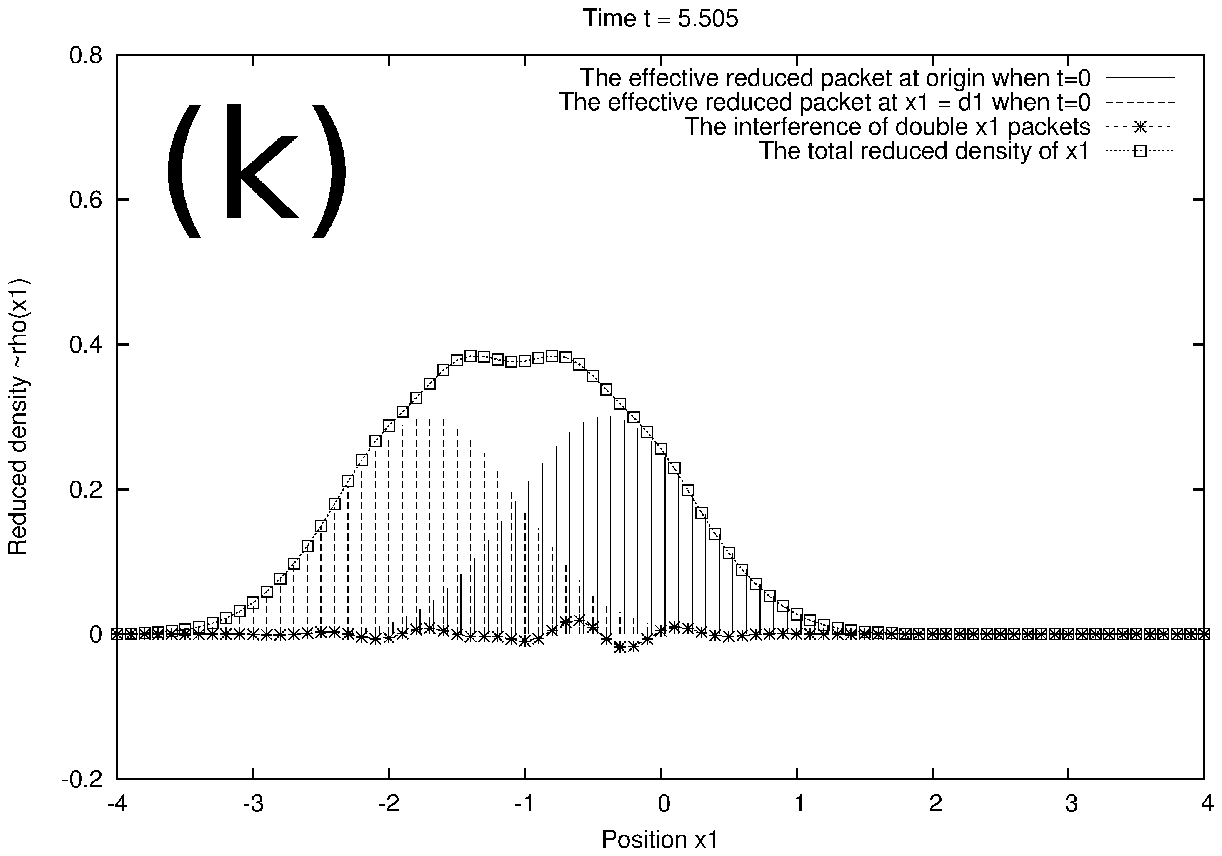}
\includegraphics[scale=.39]{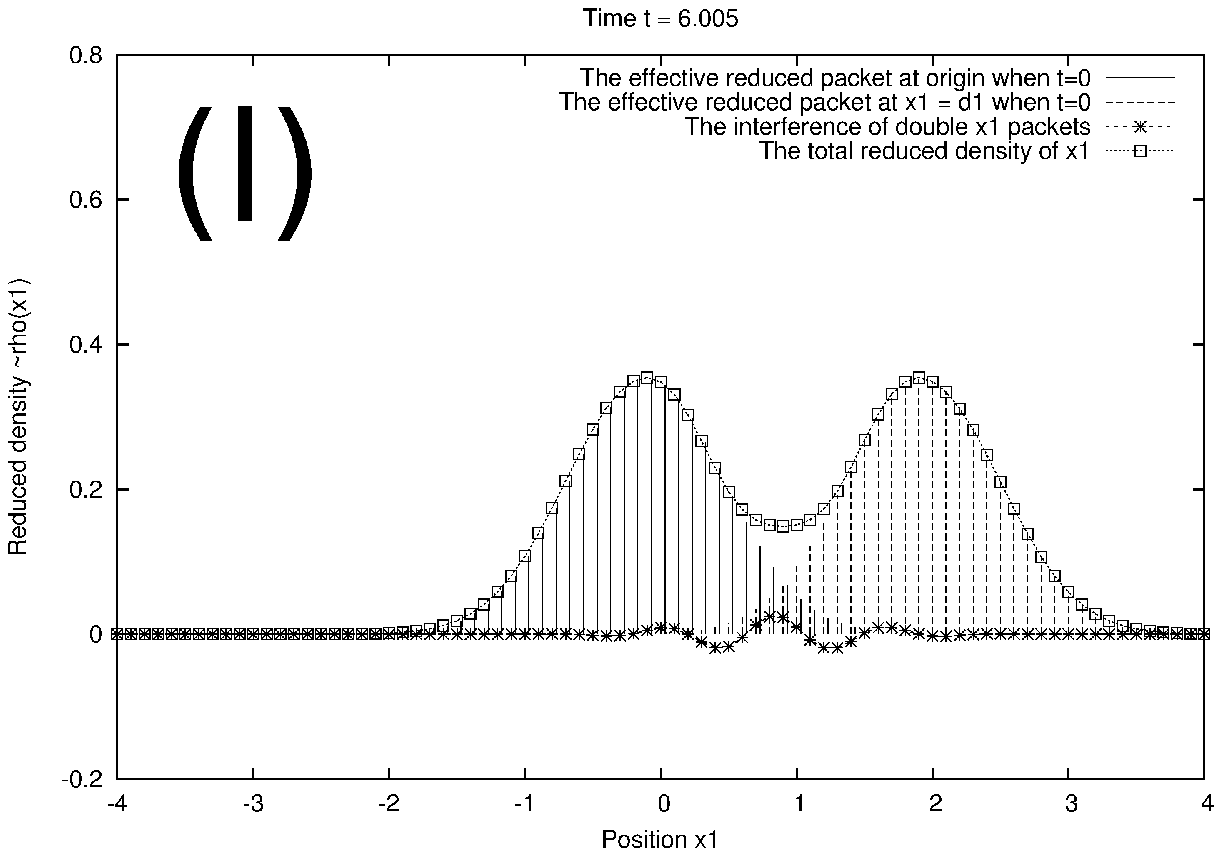}

\includegraphics[scale=.28]{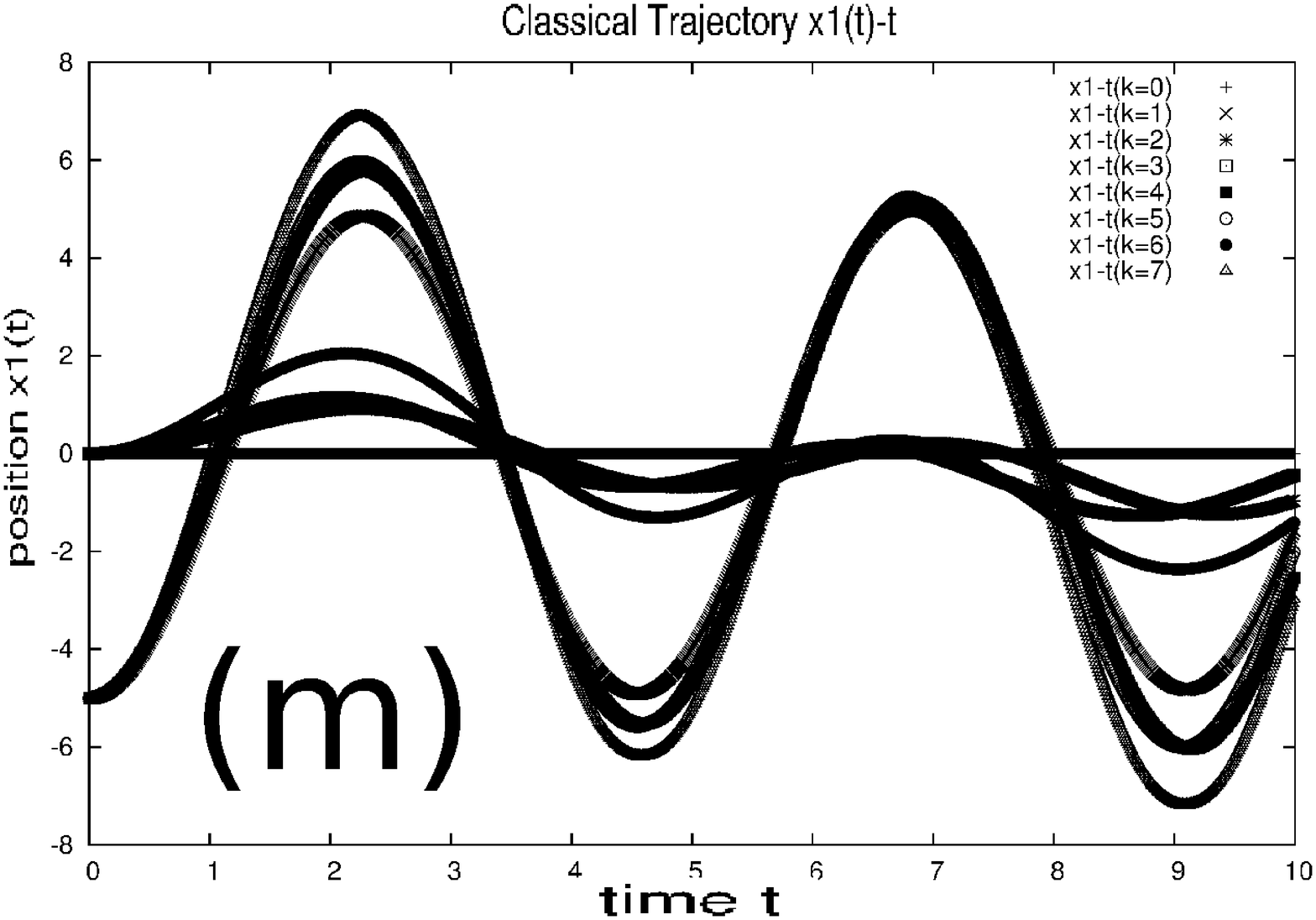}
\includegraphics[scale=.28]{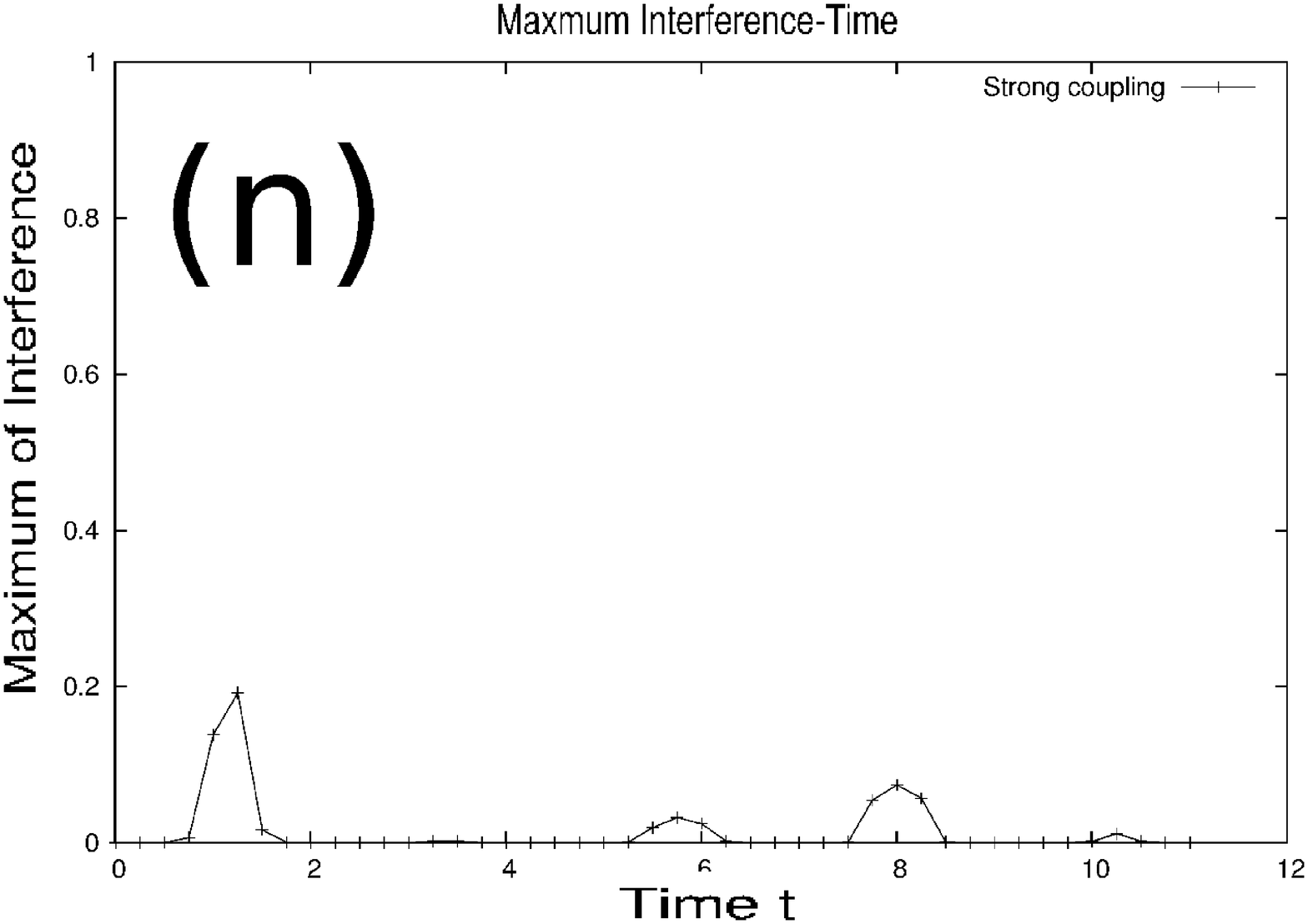}
\caption{Graphs (a)-(l) show a time evolution of the reduced dencity of particle-1 for strong coupling case.  (a) Time t=0.505, (b) t=1.005, (c) t=1.505, (d) t=2.005, (e) t=2.505, (f) t=3.005, (g) t=3.505, (h) t=4.005, (i) t=4.505, (j) t=5.005, (k) t=5.505, (l) t=6.005. And (m) shows a time evolution of classical trajectories of particle-1, \(x_1(t)\), (n) shows a time evolution of the maximum value of quantum interference, for strong coupling case.}
\label{f2}
\end{figure}

In graphs which are derived from a quantum mechanical calculation ((a)-(l) and (n) in FIG.\ref{f2}), when two Gaussian packets are crossing, interference between them is vanished or at least is damped strongly. It means decoherence arises.

Next, as you can see in a figure of classical trajectories (Image (m) in FIG.\ref{f2}), they have been branched until eight classical trajectories crossed. Then dephasing in corresponding quantum system is enough for decoherence.

%\begin{figure}[tbh]
%(a)\includegraphics[scale=.5]{trajectories_x1_b_w.eps}

%(b)\includegraphics[scale=.5]{trajectories_x1_b_s.eps}
%\caption{ Time evolution of classical trajectories of particle-1, \(x_1(t)\). The upper graph (a) is for weak coupling case. The lower graph (b) is for strong coupling case.}
%\label{f3}
%\end{figure}

\section{Conclusion}
 By this research, it is shown that not only the intersection of classical trajectories but also branchings of classical trajectories are needed for decoherence.  In other words, it was shown that interactions between a main system and environments have to make enough branches of classical trajectories of the main system for decoherence.

 Here I have to note that chaos is not need for decoherence, because this model does not include chaos. Indeed, there are a lot of worthy work about the relationship between chaos and decoherence or entanglement$^{6,7}$. And chaos seems to be good to make branchings of classical trajectories. But like this model, collecting all possible trajectories of main system corresponding to all initial conditions of all degrees of freedom which include environmental degrees of freedom, branches of classical trajectories will be reproduced without chaos or randomness.

 This discussion is a little classical aspect, so, in quantum mechanical system, what is corresponding to the collection of classical trajectories is, probably, integration procedures for environmental degrees of freedom when we get reduced density function for interested system, see Eq.(\ref{e4}). Therefore maybe we can conclude that integration procedure itself makes decoherence. Of course, in addition to it, there would be necessary condition for integrand, probability function of total Hamiltonian system. This necessary condition must be corresponding to the branchings of classical trajectories.

{\bf Acknowledgments}\\
 I would like to thank my family, Yuzuki, Miyu and Yuko Ishikawa. I would like to thank Prof.Fumihiko Sakata. I am so sorry I could not get Ph.D.  I would like to thank author of ``Miso no Keisan Buturigaku(http://www.geocities.jp/supermisosan/)''. I would like to thank researchers and technicians in NIMS Photovoltaic Materials Group. \\[5pt]
%\end{acknowledgment}
{\bf References}\\
1. W.H.Zurek, Los Alamos Science \textbf{27},2 (2002).\\
2. S.Takagi, {\em Kyoshiteki Tonneru Gensho} (Macroscopic tunneling)(IwanamiShoten,Tokyo,1997)[in Japanese].\\
3. T.Ishikawa, in {\it  Proceedings of the 12th Asia Pacific Physics Conference (APPC12),Japan,2013} edited by M. Sasao, JPS Conf. Proc.\textbf{1}, 012133 (2014).\\
4. A.O.Caldeira and A.J.Leggett, Physical Review A \textbf{31},1059 (1985). \\
5. A.O.Caldeira and A.J.Leggett, Physica A \textbf{121}, 587 (1983). \\
6. W.H.Zurek and J.P.Paz, Phys.Rev.Lett. \textbf{72}, 2508 (1994).\\
7. H.Fujisaki, T.Miyadera and A.Tanaka, Physical Review E \textbf{67}, 066201 (2003)\\
8. I.Ragnarsson and S.G.Nilsson, {\em Shapes and Shells in Nuclear Structure }(Cambridge University Press,2005)\\
9. H.Kuratsuji, {\em Genshikaku Kenkyu} \textbf{28} No.2, 3-33 (Genshikaku Danwakai,1983)
\end{document}